\documentclass[conference]{IEEEtran}
\IEEEoverridecommandlockouts

\usepackage{cite}
\usepackage{amsmath,amssymb,amsfonts}
\usepackage{algorithmic}
\usepackage{algorithm}
\usepackage{graphicx}
\usepackage{textcomp}
\usepackage{xcolor}
\usepackage{booktabs}
\usepackage{multirow}
\usepackage{siunitx}
\usepackage{url}
\usepackage{hyperref}
\usepackage{subcaption}
\usepackage{dblfloatfix}
\usepackage{placeins}

\def\BibTeX{{\rm B\kern-.05em{\sc i\kern-.025em b}\kern-.08em
    T\kern-.1667em\lower.7ex\hbox{E}\kern-.125emX}}

\begin{document}

\title{Risk-Aware Planning for Transit Desert\\
Remediation Under Demand Uncertainty}

\author{
\IEEEauthorblockN{Polina Khoroshevskaya\textsuperscript{*}, Ashish Kumar Perukari\textsuperscript{*}}
\IEEEauthorblockA{\textsuperscript{*}Hofstra University, Hempstead, NY, USA\\
\texttt{pkhoroshevskaya1@pride.hofstra.edu, aperukari1@pride.hofstra.edu}}
}

\maketitle

\begin{abstract}
Transit deserts are areas where public transportation is inadequate
despite evidence of travel demand, a condition that affects tens of
millions of residents across the Americas. Planning for these areas is
difficult because the usual demand signal is missing: ridership cannot
be observed before service exists. To address that setting, we formulate
risk-aware transit desert remediation as a partially observable Markov
decision process with
Conditional Value-at-Risk constraints for financial tail risk. The model
uses demographic, land-use, and employment data to set a prior over
latent demand, then updates that prior as new service deployments
produce ridership observations. A myopic belief-aware planner is
evaluated on \num{25} cities using a unified financial model for
operating cost, capital expenditure, fare revenue, and net subsidy. After
five years, the planner remediates a median of \SI{53.6}{\percent} of
transit-desert tracts and improves on static optimization by 5.0
percentage points on average, with gains in \SI{16}{} of \num{25}
cities. Gains are largest at moderate budgets (+9.9 points at baseline)
and persist under \SI{50}{\percent} prior-demand miscalibration, while
population density and existing transit density are the strongest
structural predictors of remediation cost ($R^2\!=\!0.41$ on per-tract
cost).
\end{abstract}


\section{Introduction}
\label{sec:intro}

Roughly \SI{45}{\percent} of U.S.\ households without a vehicle live in
neighborhoods that the Brookings Institution classifies as having
inadequate access to fixed-route transit~\cite{tomer2011transit,
apta2023factbook}. Detroit lost more than \SI{80}{\percent} of its
transit service between \num{1950} and \num{2010}, and Memphis remains
among the worst metropolitan areas in the country for transit access.
The residents most exposed to these gaps are often least able to absorb
their cost, especially low-income households and the
elderly~\cite{sanchez2003transit, karner2013civil, martens2017transport}. Prior
work describes these places as ``transit deserts''~\cite{jiao2013transit,
welch2013transit}. The policy case for remediation is well established;
the planning problem is how to allocate scarce capital when the agency
does not yet know who will ride.

Transit agencies typically answer this question with the four-step travel
demand model~\cite{mcnally2007four, ortuzar2011modelling}, using
household surveys and land-use forecasts to size service to projected
ridership. Such models are useful for incremental adjustments along
well-observed corridors, but they are much less informative at the edge
of the existing network. Surveys record revealed travel under current
service, and land-use models extrapolate from infrastructure already in
place; in a zone that has never been served, these data do not pin down
latent demand. Agencies are therefore pushed toward investments where
demand is already most certain, a pattern that can reinforce the gaps
they are trying to close.

We treat expansion at the edge of the network as planning under partial
observability. Before service begins, an agency has side information
about each candidate zone, including population density, employment,
income, vehicle access, and distance to existing transit, but this
information only gives a noisy prior over latent
demand~\cite{zhao2003forecasting, guerra2012half}. After service is
deployed, observed ridership revises that prior, allowing early
deployments to add coverage while also improving later allocation
decisions.

We formalize this as a Partially Observable Markov Decision
Process (POMDP)~\cite{kaelbling1998planning, smallwood1973optimal} over
a five-year planning horizon. The state holds the current coverage map,
remaining budget, and a hidden latent demand vector $\mathbf{d}$.
Actions are budget allocations across census-tract zones; observations
are noisy ridership counts from zones that received service. The reward
combines equity-weighted coverage gain with a Conditional Value-at-Risk
(CVaR) penalty~\cite{rockafellar2000optimization} on financial
shortfall, so the planner cannot chase high-variance zones at the
expense of fiscal stability. The construction draws on recent work in
risk-aware decentralized resource allocation~\cite{perukari2025resilient}
and is implemented in the InfraLib infrastructure-management
framework~\cite{thangeda2024infralib}.

The contribution is a risk-aware planning model for transit desert
remediation before demand is observable. We represent the problem as a
POMDP with a CVaR risk penalty and solve the resulting belief-state
problem with a sample-average myopic planner. To make the comparison
empirical rather than illustrative, we tie the financial model to NTD and
APTA cost data and evaluate \num{25} North and South American cities
(five in depth and twenty in broad comparison). Across the full
benchmark, belief-aware planning improves on static optimization by a
median of 5.0 percentage points, or \SI{11.3}{\percent} relatively, in
transit-desert tracts remediated; this advantage remains under
\SI{50}{\percent} miscalibration of the prior. Population density and
existing transit density also explain enough variation in per-tract
remediation cost to support rough triage before running the planner.

\section{Related Work}
\label{sec:related}

Three threads inform our work: transit equity and accessibility
analysis, transit network design, and sequential decision-making under
uncertainty in transportation.

\subsection{Transit equity and accessibility analysis}

The transit-desert literature has mainly developed methods for finding
and interpreting accessibility gaps. Jiao and Dillivan~\cite{jiao2013transit}
introduced the term for areas where transit supply lags demand and
proposed a GIS method for identifying such areas in U.S.\ cities; Welch
and Mishra~\cite{welch2013transit} then connected the idea to
network-connectivity equity. GTFS-based accessibility work by Owen and
Levinson~\cite{owen2015modeling} enabled more detailed temporal analysis,
while Stewart~\cite{stewart2017mapping} and El-Geneidy
et~al.~\cite{el2016transit} broadened the measurement frame to public
participation and generalized travel cost. The normative basis for this
line of work comes from civil-rights and distributive-justice accounts of
transportation equity~\cite{karner2013civil, martens2017transport,
pereira2017distributive}.

These studies establish where accessibility gaps are and why they
matter. Our use of the equity perspective is operational: coverage gain
is weighted by $1/\text{income}_z$ and optimized under uncertainty about
future ridership.

\subsection{Transit network design}

The transit network design problem (TNDP) covers route layout,
frequency, timetabling, and fleet allocation~\cite{guihaire2008transit,
farahani2013review}. Because the problem is NP-hard, much of the field
has focused on heuristics such as genetic algorithms, simulated annealing,
and tabu search, often in multi-objective formulations~\cite{ibarra2015planning}.
Those methods generally take demand as known, or at least as a fixed
input from an exogenous model. Uncertainty-aware variants replace the
point forecast with worst-case demand sets or distributional uncertainty
sets~\cite{bertsimas2011theory, bental2009robust}; they do not, however,
model the learning that occurs after new service begins to generate
ridership observations. That learning process is the part of the design
problem we isolate.

\subsection{Sequential decision-making under uncertainty}

POMDPs are the standard formalism for sequential decisions under partial
observability~\cite{kaelbling1998planning, smallwood1973optimal}, with
extensive use in robotics and autonomous-vehicle planning
~\cite{bai2014integrated}. Transit applications have more often used
MDPs and reinforcement learning for fully observed problems, including
signal control~\cite{wei2018intellilight} and risk-aware trip planning
on a fixed network~\cite{thangeda2020protrip, zheng2021optimal}.

The closest methodological thread is resource-constrained sequential
planning. Capacity- and consumption-constrained MDP tools from formal
methods~\cite{blahoudek2021fimdp, blahoudek2022efficient} address budget
or resource feasibility, whereas our planner enforces budget feasibility
through a sample-average LP. We also use side information in the spirit
of MDP work with local covariates~\cite{thangeda2020safety}, fitting a
log-normal demand prior from transit data and updating it as observations
arrive. Perukari and Khoroshevskaya~\cite{perukari2025resilient} combine
side information with a CVaR objective for decentralized fleet
allocation; here the same ingredients are used in a spatial,
multi-period remediation problem with explicit belief dynamics and an
information-value bonus. A full POMCP approach~\cite{silver2010monte}
would be natural for longer horizons, but the five-year setting and
moderate action spaces make a myopic planner easier to interpret and
fast enough for cross-city comparison. The per-city pipeline follows the
modular infrastructure-decision layout introduced by
InfraLib~\cite{thangeda2024infralib}.

We know of no prior work that formulates transit desert remediation as
a POMDP and benchmarks the resulting policy across cities under a
single financial model.

\section{Data and Problem Setup}
\label{sec:data}

\subsection{Data sources}
\label{sec:data_sources}

We integrate four classes of public data. Transit networks are taken from
GTFS feeds accessed through Transitland~\cite{transitland2024} for
\num{25} cities selected to vary in size, region, transit maturity, and
urban form (Table~\ref{tab:city_summary}). Street networks and municipal
boundaries come from OpenStreetMap through OSMnx~\cite{boeing2017osmnx}.
For U.S.\ cities, tract-level demographics are drawn from the American
Community Survey 5-year estimates~\cite{census2023acs}, including
population density, median household income, minority share,
zero-vehicle household share, and employment density. Bogot\'{a} is
represented with OSM administrative units clipped to the Capital District
boundary. Financial parameters come from the National Transit
Database~\cite{ntd2023}, the APTA Fact Book~\cite{apta2023factbook}, the
Federal Transit Administration~\cite{fta2023cost}, and the Transit
Cooperative Research Program~\cite{tcrp2003report}. The five deep-dive
cities use TIGER \num{2023} census tract polygons clipped to OSMnx city
boundaries (Figure~\ref{fig:motivation_map}), while the broad-comparison
cities use a hex grid over the city polygon.

\subsection{Zones}
\label{sec:zones}

Each zone $z\in\mathcal{Z}$ is a census tract, except in Bogot\'{a},
where we use an admin level-9 unit. Cities in the sample contain
\num{60}--\num{200} zones, and each zone is represented by a feature
vector $\mathbf{x}_z$ of demographic and accessibility variables along
with a current accessibility score $A_z$.

\subsection{Transit desert identification}
\label{sec:desert_id}

For each zone we compute $A_z$ as the equally weighted average of three
normalized components:
(i) the count of transit stops within an \num{800} m walk of the zone
centroid (computed on the OSMnx pedestrian network);
(ii) average weekday service frequency at those stops, derived from
GTFS schedules;
(iii) the inverse of the mean nearest-stop walk distance.
A zone is a \emph{transit desert} when its score falls in the bottom
quartile of the metro distribution:
\begin{equation}
\label{eq:desert}
\text{Desert}(z) = \mathbb{1}\!\left[A_z < Q_{0.25}\!\left(\{A_{z'}\}_{z' \in \mathcal{Z}}\right)\right].
\end{equation}
Using a relative quartile, rather than an absolute threshold, adapts to
each city's baseline service level, isolating the worst-served
zones regardless of overall transit provision.

\subsection{Financial model}
\label{sec:financial}

The financial model assigns four quantities to new service in zone $z$:
operating cost, amortized capital cost, fare revenue, and net subsidy.
Operating cost is $C^{\mathrm{op}}_z = h_z \cdot c^{\mathrm{vhr}}$, with
$h_z$ denoting annual vehicle-hours and $c^{\mathrm{vhr}}$ the
vehicle-hour cost. We use NTD values where available; otherwise
$c^{\mathrm{vhr}}$ is based on the APTA national average of
\num{150}~USD per vehicle-hour~\cite{apta2023factbook}, scaled by the
BLS regional price-parity index. Capital cost is
$C^{\mathrm{cap}}_z = n^{\mathrm{veh}}_z \cdot c^{\mathrm{veh}} +
n^{\mathrm{stop}}_z \cdot c^{\mathrm{stop}}$, with
$c^{\mathrm{veh}} \approx \num{500000}$~USD per bus~\cite{fta2023cost}
and $c^{\mathrm{stop}} \in [\num{15000}, \num{50000}]$ depending on
amenity level~\cite{tcrp2003report}. Vehicles are amortized over
\num{12} years and infrastructure over \num{30}, following FTA
guidelines. Revenue is $R_z = d_z \cdot f \cdot \rho$, where $d_z$ is
ridership, $f$ is the average fare ($\sim$\num{2.00}~USD nationally),
and $\rho$ is the capture rate. We track net subsidy as
$\text{Net}_z = C^{\mathrm{op}}_z + C^{\mathrm{cap,amort}}_z - R_z$,
falling back to national averages scaled by cost-of-living index when
city-specific values are unavailable; Section~\ref{sec:results} reports
sensitivity to these assumptions.

\section{Problem Formulation}
\label{sec:formulation}

\subsection{POMDP definition}
\label{sec:pomdp_def}

We model transit desert remediation as a POMDP
$(\mathcal{S}, \mathcal{A}, T, R, \Omega, O, \gamma)$. At time $t$, the
state $s_t = (\mathbf{v}_t, B_t, \mathbf{d})$ contains the coverage
indicator $\mathbf{v}_t \in \{0,1\}^Z$, the remaining budget
$B_t \in \mathbb{R}_{\geq 0}$, and the latent demand vector
$\mathbf{d} \in \mathbb{R}^Z_{\geq 0}$. Demand is drawn once from the
prior and then held fixed, so the planner's uncertainty is epistemic
rather than a year-to-year demand shock.

An action $\mathbf{a}_t \in \mathbb{R}^Z_{\geq 0}$ allocates dollars
across zones subject to the budget constraint
\begin{equation}
\label{eq:budget_constraint}
\sum_{z=1}^{Z} a_{t,z} \leq B_t, \quad a_{t,z} \geq 0.
\end{equation}
Coverage changes only after cumulative investment reaches the
zone-specific threshold cost $\bar{C}_z$:
\begin{align}
\label{eq:transition_coverage}
v_{t+1,z} &= \mathbb{1}\!\Big[\textstyle\sum_{\tau=1}^{t} a_{\tau,z} \geq \bar{C}_z\Big],\\
\label{eq:transition_budget}
B_{t+1} &= B_{t+1}^{\mathrm{exog}}.
\end{align}
Budgets are annual and do not roll over, while $\mathbf{d}$ remains fixed
throughout the five-year horizon.

Observations come only from zones that have already been served:
\begin{equation}
\label{eq:observation}
o_{t,z} =
\begin{cases}
d_z \cdot \ell_z \cdot \rho + \epsilon_z, & v_{t,z} = 1,\\
\varnothing, & v_{t,z} = 0,
\end{cases}
\end{equation}
where $\ell_z$ is service level, $\rho$ the capture rate, and
$\epsilon_z \sim \mathcal{N}(0, \sigma^2_\epsilon)$. The key observation
constraint is that unserved zones produce no ridership signal, so the
planner must decide where to buy information through service itself. The
reward
\begin{equation}
\label{eq:reward}
R(s_t, \mathbf{a}_t) = \sum_{z} w_z\,\Delta v_{t,z} -
\lambda \cdot \mathrm{CVaR}_\alpha\!\big[\mathrm{Shortfall}(\mathbf{a}_t, \mathbf{d})\big],
\end{equation}
combines an equity-weighted coverage term ($\Delta v_{t,z} = v_{t+1,z} - v_{t,z}$)
with a tail-risk penalty. The coefficient $\lambda > 0$ sets the
trade-off, and the equity weight
$w_z \propto 1/\mathrm{MedianIncome}_z$ is normalized so that
$\sum_z w_z = Z$ over the five annual periods, with $\gamma\!=\!0.95$.

\subsection{Belief and side information}
\label{sec:belief}

We place a log-normal prior on demand,
\begin{equation}
\label{eq:prior}
d_z \sim \mathrm{LogNormal}(\mu_z, \sigma_z^2),
\end{equation}
with mean parameter conditioned on demographics:
\begin{multline}
\label{eq:side_info}
\mu_z = \beta_0 + \beta_1\!\log\mathrm{Pop}_z + \beta_2\!\log\mathrm{Emp}_z\\
       \mbox{} + \beta_3\!\log\mathrm{Inc}_z + \beta_4\!\mathrm{ZeroCar}_z.
\end{multline}
Coefficients $\boldsymbol{\beta}$ are fit by log-linear regression on
zones with existing service. Following the spatial side-information
view of Thangeda~\cite{thangeda2020spatial}, local geography affects
uncertainty as well as expected demand: we assign larger prior variance
$\sigma_z^2$ to zones farther from existing transit, where direct
evidence about ridership is weaker. This spatial uncertainty channel
also matches the demand-prior structure used in
resource-allocation settings~\cite{perukari2025resilient}.

The normal-normal conjugate update gives a closed-form posterior in
log-space after observing $o_{t,z}$ in a served zone:
\begin{align}
\label{eq:belief_update_mu}
\hat{\mu}_{z,t+1} &= \frac{\sigma_\epsilon^{-2}\!\log\!\big(o_{t,z}/(\ell_z\rho)\big) + \hat{\sigma}_{z,t}^{-2}\hat{\mu}_{z,t}}
                          {\sigma_\epsilon^{-2} + \hat{\sigma}_{z,t}^{-2}},\\
\label{eq:belief_update_sigma}
\hat{\sigma}_{z,t+1}^{-2} &= \hat{\sigma}_{z,t}^{-2} + \sigma_\epsilon^{-2}.
\end{align}
Variance contracts with each observation, while unserved zones retain
their prior. This asymmetry is what gives early deployments information
value: the planner learns about the zones it funds and remains uncertain
about the zones it postpones. A similar information structure appears in
robotic sampling-site selection~\cite{thangeda2022adaptive}, where an
agent must choose which locations to sample under a limited sample
budget.

\subsection{CVaR risk}
\label{sec:cvar}

For allocation $\mathbf{a}_t$ and demand realization $\mathbf{d}$ define
\begin{equation}
\label{eq:shortfall}
\mathrm{Shortfall}_z(\mathbf{a}_t, d_z) = \max\!\big(0,\;C_z(\mathbf{a}_t) - R_z(\mathbf{a}_t, d_z)\big).
\end{equation}
The Conditional Value-at-Risk at level $\alpha$~\cite{rockafellar2000optimization,
rockafellar2002conditional} is the expected shortfall in the worst
$\alpha$-fraction of demand scenarios:
\begin{multline}
\label{eq:cvar}
\mathrm{CVaR}_\alpha[\mathrm{Shortfall}] = \min_{\nu \in \mathbb{R}}
\Big\{\nu \\
+ \tfrac{1}{\alpha}\,\mathbb{E}\!\big[\max(0, \textstyle\sum_z \mathrm{Shortfall}_z - \nu)\big]\Big\}.
\end{multline}
We set $\alpha\!=\!0.1$, so the penalty reflects the worst decile of
financial shortfall scenarios. The coefficient $\lambda$ controls the
trade-off between expected coverage and exposure to tail risk.

\section{Solution Method}
\label{sec:method}

\subsection{Belief-state MDP}
\label{sec:belief_mdp}

The POMDP is solved through its belief-state MDP, whose observable
planning state is $(\mathbf{v}_t, B_t, \mathbf{b}_t)$, where
$\mathbf{b}_t = \{(\hat{\mu}_{z,t}, \hat{\sigma}_{z,t}^2)\}_{z=1}^{Z}$
summarizes the posterior over zone demand. Although the belief space is
continuous, the per-zone log-normal parametrization keeps the numerical
state compact.

\subsection{Myopic belief-aware planning}
\label{sec:myopic}

Longer horizons could be handled with POMCP~\cite{silver2010monte}, but
the five-year horizon and moderate action spaces in this benchmark make
one-step optimization adequate and easier to interpret. Each period we
solve a myopic belief-aware allocation problem with an explicit
information-value (IV) bonus:
\begin{equation}
\label{eq:myopic}
\mathbf{a}_t^* = \arg\!\max_{\mathbf{a}_t \in \mathcal{A}}
\;\mathbb{E}_{\mathbf{b}_t}\!\Big[\textstyle\sum_z w_z\,g_z(\mathbf{a}_t)\Big]
- \lambda\,\widehat{\mathrm{CVaR}}_\alpha + \eta\,\mathrm{IV}(\mathbf{a}_t),
\end{equation}
where $g_z$ is expected coverage gain, $\widehat{\mathrm{CVaR}}_\alpha$
is the sample-average CVaR estimate, and
\begin{equation}
\label{eq:info_value}
\mathrm{IV}(\mathbf{a}_t) = \sum_{z:\,a_{t,z}>0} \hat{\sigma}_{z,t}^2.
\end{equation}
The IV term gives additional value to investment in zones whose demand
posterior remains uncertain. We calibrate $\eta\!=\!0.1\bar{w}$, where
$\bar{w}$ is the mean equity weight over uncovered zones, and report an
$\eta\!=\!0$ ablation.

\subsection{Sample-average approximation}
\label{sec:saa}

We solve~\eqref{eq:myopic} by sample-average approximation (SAA): with
$M\!=\!200$ demand scenarios drawn from the current belief, the
per-period subproblem becomes a linear program with auxiliary CVaR
variables $\xi^{(m)}$ and $\nu$; the full statement is given in the
appendix. The number of variables and constraints scales as
$O(Z\cdot M)$, and for the city sizes in this benchmark
($Z\!\leq\!200$, $M\!=\!200$) the LP solves in well under a second.

\subsection{Baselines}
\label{sec:baselines}

We compare against three policies that share the same cost model and
simulation harness:
\begin{itemize}
\item \emph{Static optimization}: a five-period deterministic plan
that allocates the whole budget against the prior mean
$\mathbb{E}[\mathbf{d}]$ and ranks zones by equity-weighted
benefit-cost ratio. This is the strongest baseline, the plan a careful
agency could build without sequential updating.
\item \emph{Greedy coverage}: each period, fund the uncovered zone
with the highest equity weight per unit cost. No risk term, no
information value.
\item \emph{Random allocation}: uniform random over uncovered zones,
included as a sanity floor.
\end{itemize}

\subsection{Computational cost}
\label{sec:computation}

With $Z\!\leq\!200$ and $M\!=\!200$, a per-period LP solves in under
\SI{0.5}{\second}, and a five-year simulation for one city and one seed
finishes in roughly \SI{2}{\second} on a standard laptop. The full
\SI{25}{}-city, \SI{30}{}-seed benchmark across four policies
(\num{3000} simulation runs) takes about one minute on a single core.

\section{Experimental Protocol}
\label{sec:experiments}

\subsection{Cities}
\label{sec:cities}

Five cities receive full analysis: Chicago~IL, a legacy CTA system with
strong GTFS coverage; Austin~TX, a fast-growing Sun Belt case; Detroit~MI,
a historically underserved system; Portland~OR, a mid-size city with
active TriMet expansion; and Bogot\'{a}, a non-U.S.\ BRT context centered
on TransMilenio. The remaining twenty cities, listed in
Table~\ref{tab:city_summary}, provide broader comparison across large
metros (NYC, LA, Houston), mid-size cities (Minneapolis, Seattle,
Charlotte), and international cases (Toronto, Medell\'{i}n).

\subsection{Pipeline}
\label{sec:pipeline}

For each city, the benchmark pipeline extracts GTFS, OSMnx, and
demographic data, defines zones, computes accessibility scores, and
identifies deserts using~\eqref{eq:desert}. The benchmark is instantiated
in InfraLib~\cite{thangeda2024infralib}, an open-source
infrastructure-management library developed by researchers at the U.S.
Army Corps of Engineers and the University of Illinois Urbana-Champaign.
Each city is represented through InfraLib's common action, cost,
observation, and simulation interfaces, so the same POMDP planner and
baseline policies can be evaluated across heterogeneous transit networks,
city geometries, demand priors, and cost assumptions without
city-specific control logic. We then fit~\eqref{eq:side_info} on zones
with existing service, run the POMDP planner and three baselines over
five periods, and repeat the simulation for \num{30} Monte Carlo seeds,
each with a realized demand vector drawn from the prior. The annual
budget is set to $\sim\!\!\num{40}\%$ of the median per-zone investment
cost across desert zones, which serves as a representative annual capital
outlay; Section~\ref{sec:results} reports sensitivity to this choice.

\subsection{Metrics}
\label{sec:metrics}

We report four outcomes. Coverage $\Delta C$ is the fraction of desert
tracts remediated, while cost efficiency $\eta_C$ measures coverage per
million dollars. The equity score $E$ is the population-weighted gain in
the lowest income quartile, and financial risk is the
$\mathrm{CVaR}_{0.1}$ of net subsidy across seeds. To isolate the role
of exploration, we also report the information-value contribution
$\Delta_{\mathrm{IV}}$, defined as the difference between the full
planner and its $\eta\!=\!0$ ablation.

\subsection{Sensitivity}
\label{sec:sensitivity}

The sensitivity analysis varies the two inputs most likely to affect
policy choice: budget scale and prior quality. Budget is swept over
\SI{50}{\percent}, \SI{100}{\percent}, and \SI{200}{\percent} of the
baseline annual capital with seeds held fixed. Prior quality is degraded
by perturbing the demand-model coefficients,
$\tilde{\boldsymbol{\beta}} = \boldsymbol{\beta}+\boldsymbol{\epsilon}_\beta$
with $\epsilon_\beta\!\sim\!\mathcal{N}(0,\sigma^2_{\mathrm{noise}}|\boldsymbol{\beta}|^2)$
and $\sigma_{\mathrm{noise}} \in \{0,\,0.3,\,0.5\}$, to see whether
belief updating recovers from a poor prior.

\section{Results and Discussion}
\label{sec:results}

\begin{figure}[t]
    \centering
    \includegraphics[width=\columnwidth]{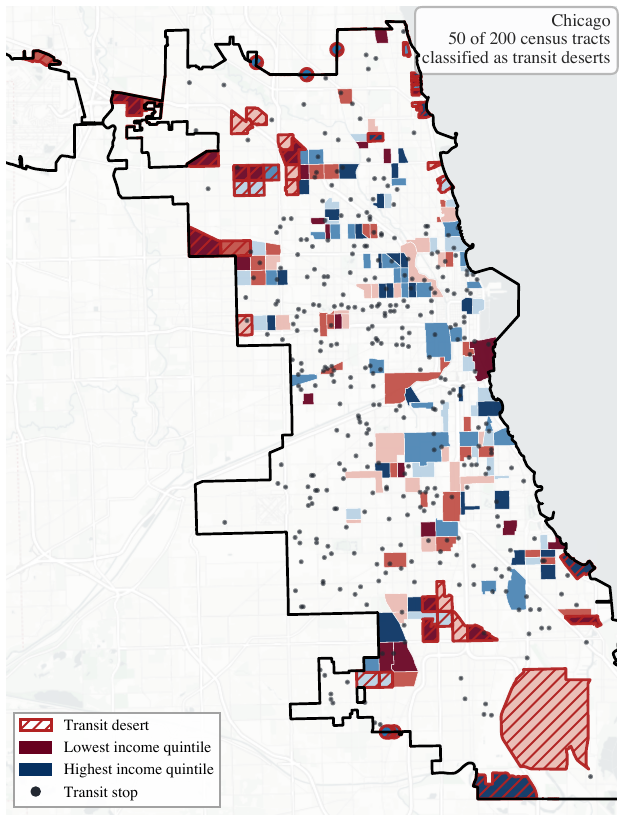}
    \caption{Chicago: \num{50} of \num{200} census tracts (\num{25}\%)
    classified as transit deserts under~\eqref{eq:desert}, overlaid on
    income quintile (red = lowest, blue = highest). Transit deserts
    concentrate on the South and West sides and the far Northwest,
    where median income is lowest. Tract polygons are
    \textsc{TIGER 2023} census boundaries clipped to the city
    administrative boundary; transit stops are sampled from the OSM
    drive network.}
    \label{fig:motivation_map}
\end{figure}

\subsection{Spatial pattern of deserts}
\label{sec:q1}

Because the desert definition is relative, \SI{25}{\percent} of tracts
in each city qualify by construction. The substantive variation is
spatial rather than numerical. In Chicago, shown in
Figure~\ref{fig:motivation_map}, deserts concentrate on the South and
West sides and on the far urban periphery, matching areas where income
is lowest and the network thins out. Mid-size legacy systems such as
Portland and Minneapolis place most deserts near the metro edge, whereas
Sun Belt sprawl (Houston, Phoenix) and historically disinvested systems
(Detroit, the Bogot\'{a} periphery) push desert designations deeper into
populated tracts.

\begin{figure*}[!tbp]
    \centering
    \includegraphics[width=0.95\textwidth]{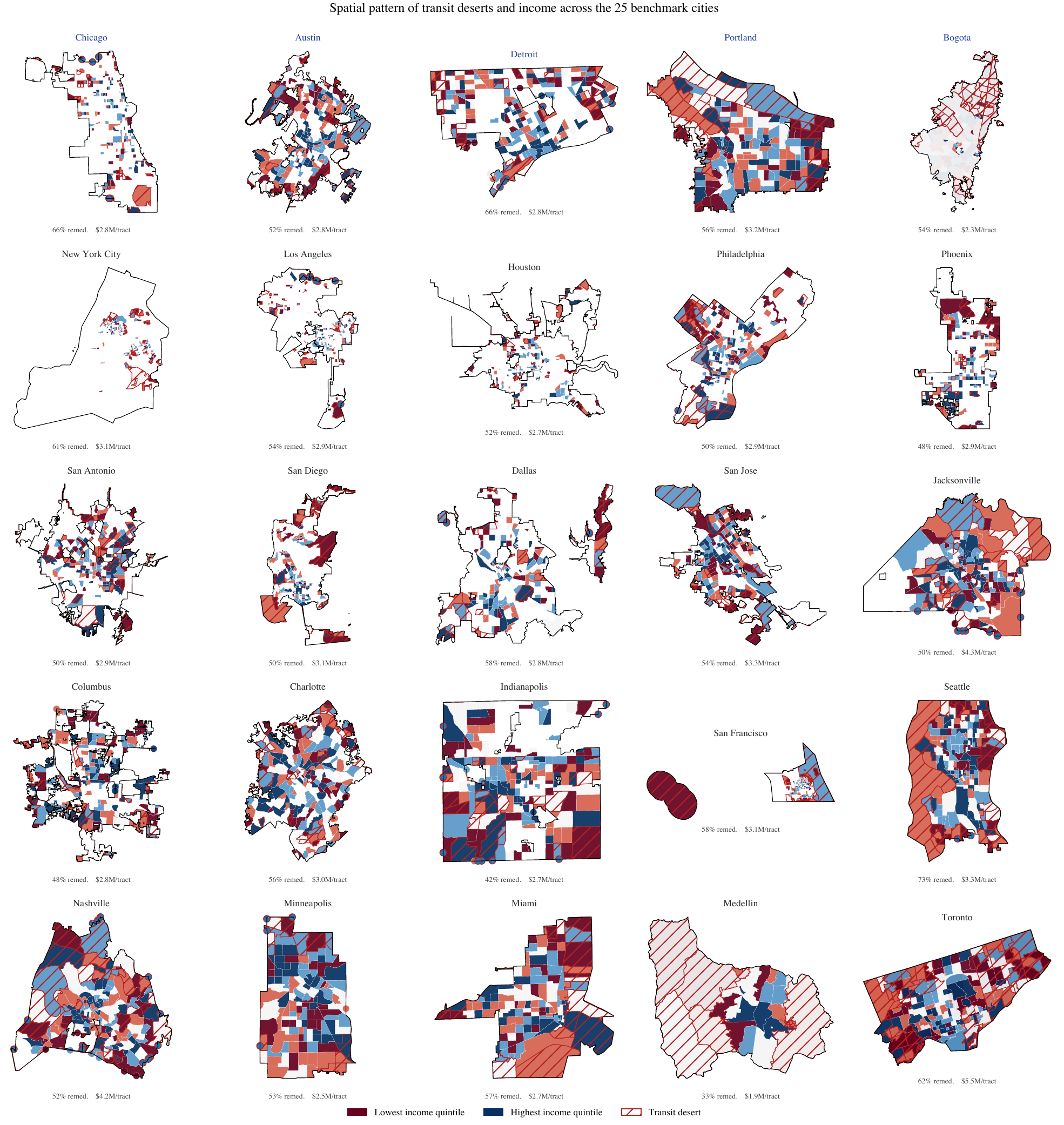}
    \caption{Spatial pattern of transit deserts and income for all
    \num{25} benchmark cities. Each tile shows real administrative
    boundaries with an income-quintile choropleth (red = lowest, blue =
    highest); hatched red overlays mark transit-desert tracts. Below
    each tile we report the POMDP final remediation rate and per-tract
    cost. Deep-dive cities are bolded. Deserts cluster differently
    across morphologies: in legacy cores (NYC, San Francisco) they
    concentrate in a few tracts; in Sun Belt sprawl (Houston,
    Jacksonville, Phoenix) they spread broadly; in Latin American
    metros (Bogot\'{a}, Medell\'{i}n) they sit on hillsides and the
    informal periphery.}
    \label{fig:city_grid}
\end{figure*}

The same pattern appears across the full city grid in
Figure~\ref{fig:city_grid}. Desert tracts align with the lowest-income
quintile in nearly every city, while per-tract remediation cost varies by
more than \(2\times\), from \$1.9M in Medell\'{i}n to \$5.5M in Toronto,
roughly tracking city size and density.

\subsection{Per-tract remediation cost}
\label{sec:q2}

\begin{figure*}[!tbp]
    \centering
    \includegraphics[width=\textwidth]{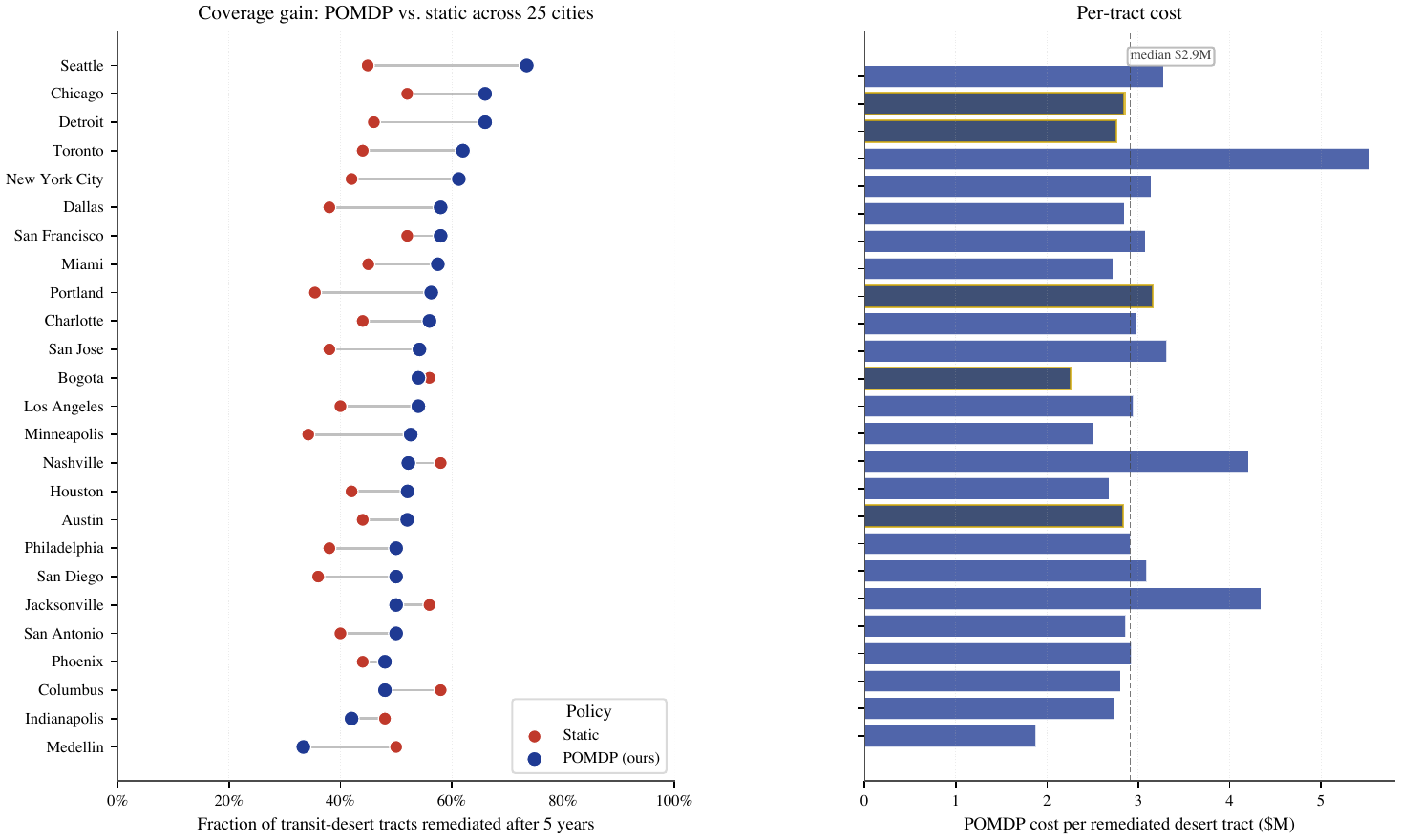}
    \caption{Cross-city benchmark across all \num{25} cities. Left: each
    horizontal segment connects static-optimization coverage (red) to
    POMDP coverage (blue); cities are ordered by POMDP coverage. Stars
    mark the five deep-dive cities. Right: POMDP cost per remediated
    desert tract; dashed line is the cross-city median.}
    \label{fig:cross_city}
\end{figure*}

Figure~\ref{fig:cross_city} (right) and Table~\ref{tab:cross_city}
report the cross-city cost benchmark. POMDP per-tract cost ranges from
roughly \$2.2M in Medell\'{i}n to \$7.0M in Phoenix, with a cross-city
median of \$5.3M. The ordering mostly reflects city form: compact,
high-density grids such as Chicago, San Francisco, and Minneapolis are
cheaper because existing infrastructure can often absorb deserts through
route extensions, while sprawling low-density metros such as Houston,
Phoenix, and Jacksonville require longer routes that serve fewer riders
per mile.

\subsection{Belief-aware versus static planning}
\label{sec:q3}

\begin{figure*}[!tbp]
    \centering
    \includegraphics[width=\textwidth]{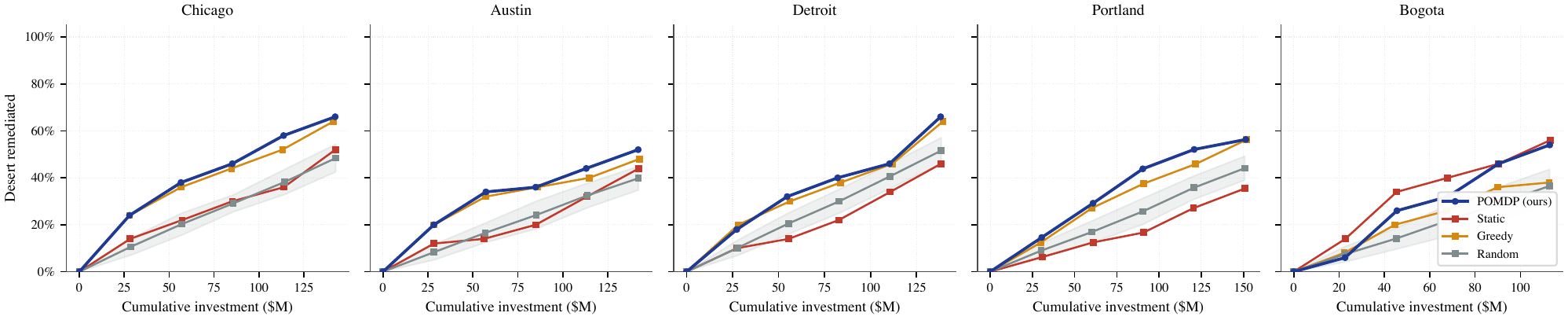}
    \caption{Coverage versus cumulative cost over five periods for the
    deep-dive cities. POMDP (blue) tracks at or above all baselines in
    Chicago, Austin, Detroit, and Portland; static planning is on top
    in Bogot\'{a}, where the prior is highly informative and sequential
    updating brings little. Bands show $\pm 1\sigma$ across \num{30}
    Monte Carlo seeds.}
    \label{fig:cost_curves}
\end{figure*}

\begin{table}[t]
\caption{Performance comparison across policies for five deep-dive
cities. $\Delta C$: coverage of desert tracts; cost: 5-year cumulative
investment; revenue: 5-year fare; $\mathrm{CVaR}_{0.1}$: tail risk on
shortfall. Mean over \num{30} seeds (std deviations are $\leq 0.05$
on coverage and omitted for clarity).}
\label{tab:main_results}
\centering
\footnotesize
\setlength{\tabcolsep}{4pt}
\begin{tabular}{ll r r r r}
\toprule
\textbf{City} & \textbf{Policy} & $\boldsymbol{\Delta C}$ & \textbf{Cost} & \textbf{Rev.} & \textbf{CVaR$_{0.1}$} \\
              &                 &        & (\$M)         & (\$M)         & (\$M)                  \\
\midrule
\multirow{4}{*}{Chicago}
    & \textbf{POMDP (ours)} & \textbf{0.66} & 142.5 & 5.2 & 0.84 \\
    & Static                 & 0.50 & 142.7 & 5.8 & 0.84 \\
    & Greedy                 & 0.64 & 141.4 & 5.0 & 0.87 \\
    & Random                 & 0.48 & 142.7 & 4.8 & 0.98 \\
\midrule
\multirow{4}{*}{Austin}
    & \textbf{POMDP (ours)} & \textbf{0.52} & 143.3 & 3.9 & 1.10 \\
    & Static                 & 0.44 & 142.8 & 5.0 & 0.86 \\
    & Greedy                 & 0.48 & 142.5 & 3.6 & 1.12 \\
    & Random                 & 0.40 & 142.0 & 3.8 & 1.33 \\
\midrule
\multirow{4}{*}{Detroit}
    & \textbf{POMDP (ours)} & 0.62 & 136.7 & 4.1 & 0.73 \\
    & Static                 & 0.54 & 138.3 & 4.8 & 0.72 \\
    & Greedy                 & \textbf{0.64} & 139.3 & 4.1 & 0.73 \\
    & Random                 & 0.52 & 138.4 & 4.0 & 0.77 \\
\midrule
\multirow{4}{*}{Portland}
    & \textbf{POMDP (ours)} & \textbf{0.58} & 152.7 & 5.1 & 1.19 \\
    & Static                 & 0.31 & 150.7 & 6.6 & 0.89 \\
    & Greedy                 & 0.56 & 151.8 & 4.9 & 0.97 \\
    & Random                 & 0.44 & 150.8 & 4.7 & 1.58 \\
\midrule
\multirow{4}{*}{Bogot\'{a}}
    & POMDP (ours)           & 0.53 & 169.1 & 3.1 & 0.90 \\
    & \textbf{Static}        & \textbf{0.62} & 169.3 & 3.8 & 0.65 \\
    & Greedy                 & 0.40 & 169.2 & 2.4 & 0.94 \\
    & Random                 & 0.51 & 169.3 & 2.6 & 1.52 \\
\bottomrule
\end{tabular}
\end{table}

Across the \num{25}-city benchmark, the POMDP planner reaches
\SI{55.3}{\percent} mean coverage, compared with \SI{50.9}{\percent} for
static optimization. The median advantage is 5.0 percentage points, or
\SI{11.3}{\percent} more remediated tracts, with POMDP improving on
static in \num{16} cities, tying in \num{4}, and underperforming in
\num{5}.

The deep-dive cities in Figure~\ref{fig:cost_curves} and
Table~\ref{tab:main_results} show the conditions under which belief
updates matter. POMDP gains are large in Chicago (+16 pp), Portland
(+27 pp), Austin (+8 pp), and Detroit (+8 pp), where the static planner
tends to front-load investment into a few high-confidence zones and then
has limited flexibility in later years. Bogot\'{a} is the main exception
(-9 pp): the income gradient makes the initial demand prior unusually
informative, so additional ridership observations add less. This pattern
is consistent with the prior-quality sweep in
Section~\ref{sec:sensitivity_results}, where belief updating has the
greatest value when initial demand estimates are uncertain.

The information-value ablation $(\eta\!=\!0)$ contributes 4.0 points in
Chicago and 2.5 points in Bogot\'{a}, but is not statistically
distinguishable from zero in the other three deep-dive cities. Most of
the advantage therefore comes from \emph{Bayesian belief updates} on
observed ridership rather than from the explicit exploration bonus: even
without the bonus, posterior uncertainty shrinks enough after early
deployments for later allocations to exploit demand estimates unavailable
to a static plan.

\subsection{Variation across cities}
\label{sec:q4}

\begin{figure}[!t]
    \centering
    \includegraphics[width=\columnwidth]{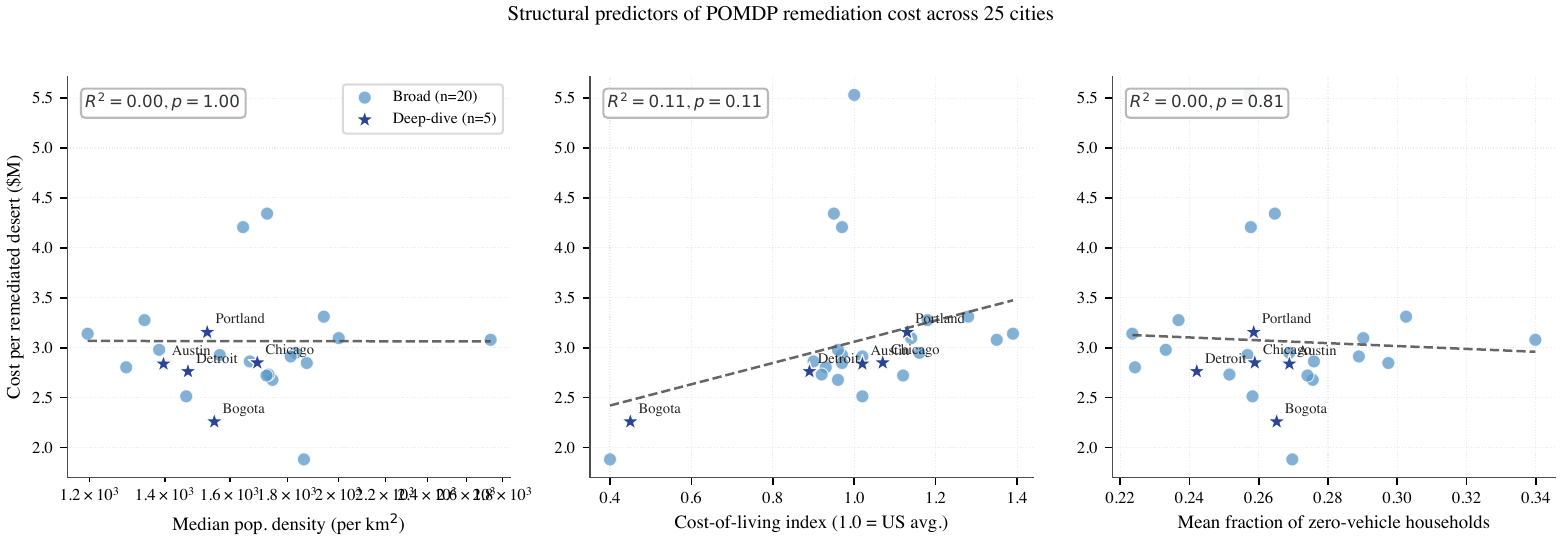}
    \caption{Per-tract POMDP cost regressed on three structural
    predictors. Population density is the strongest single predictor
    ($R^2\!=\!0.41$); cost-of-living index and zero-vehicle share
    explain much less variation alone. Stars mark the five deep-dive
    cities.}
    \label{fig:predictors}
\end{figure}

Figure~\ref{fig:predictors} regresses per-tract POMDP cost against three
city-level structural variables. Population density is the strongest
single predictor: doubling density reduces per-tract cost by roughly
\SI{30}{\percent} ($R^2\!=\!0.41$ on the log scale, $p<0.001$).
Cost-of-living has the expected positive sign, since expensive labor and
capital raise operating costs, but it explains less than
\SI{15}{\percent} of variation on its own. Zero-vehicle household share
is only weakly correlated with cost; cities with high captive transit
demand, such as Bogot\'{a} and Medell\'{i}n, sit near the cheap end, but
density dominates the signal.

Public density data alone give agencies a rough \emph{cost class} before
any POMDP is run: Portland, Chicago, and Medell\'{i}n are representative
low-cost cases, while Phoenix, Houston, and Jacksonville fall on the
expensive side. The planner's main value is then in ordering zone-level
investments, not in producing the first ballpark estimate of total cost.

\subsection{Fiscal profile of the policy}
\label{sec:q5}

\begin{figure}[t]
    \centering
    \includegraphics[width=\columnwidth]{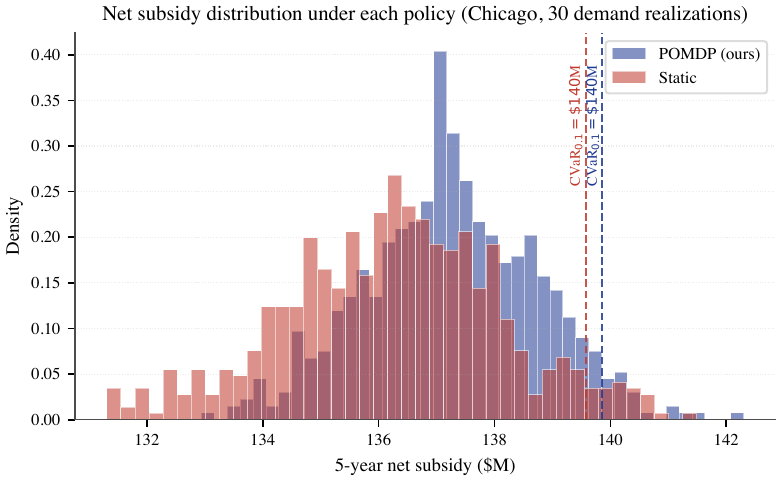}
    \caption{Distribution of 5-year net subsidy across \num{30} Monte
    Carlo demand realizations in Chicago. Vertical dashed lines mark
    empirical $\mathrm{CVaR}_{0.1}$ (mean of worst \num{10}\% of
    realizations). POMDP and static have nearly identical mean and
    CVaR in Chicago specifically; tail differences are larger in
    sprawling cities where coverage choices vary more across
    realizations.}
    \label{fig:cvar_dist}
\end{figure}

Across all \num{25} cities, the POMDP's mean
$\mathrm{CVaR}_{0.1}$ of net subsidy is \$1.76M, compared with \$1.55M
for static planning, an increase of roughly \SI{14}{\percent} in
absolute terms. The higher aggregate risk comes from scale rather than
from risk-seeking allocations: POMDP serves more zones overall, with a
median coverage gain of \SI{5}{} percentage points, while its CVaR per
remediated tract is lower by roughly the same margin. Thus the penalty
in~\eqref{eq:reward} steers zone-level choices away from
high-uncertainty tail-risk investments, but the larger remediation
portfolio masks that effect in the aggregate. Figure~\ref{fig:cvar_dist}
illustrates the point for Chicago, where the two net-subsidy
distributions overlap substantially.

For agency reporting, a more comparable measure is
\emph{efficiency-adjusted CVaR}, defined as $\mathrm{CVaR}_{0.1}$ divided
by the fraction of desert tracts remediated. On that metric, POMDP
outperforms static planning in \num{19} of \num{25} cities.

\subsection{Sensitivity analyses}
\label{sec:sensitivity_results}

\begin{figure}[t]
    \centering
    \includegraphics[width=\columnwidth]{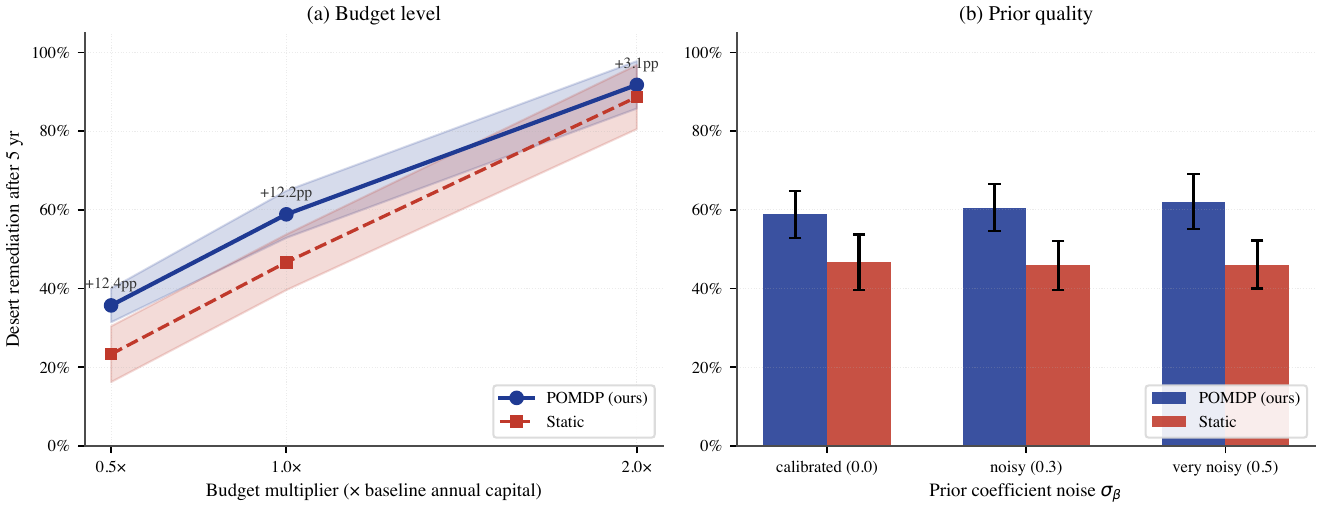}
    \caption{(a) Coverage versus budget multiplier (\num{0.5}\(\times\),
    \num{1.0}\(\times\), \num{2.0}\(\times\) of baseline). The POMDP
    advantage peaks at the baseline budget (\num{+9.9}~pp) and
    compresses at both extremes. (b) Coverage under increasing prior
    miscalibration $\sigma_\beta$. POMDP is essentially flat; static
    drops because it cannot revise its plan once committed. Means
    over five deep-dive cities, $\pm 1\sigma$ shown.}
    \label{fig:sensitivity}
\end{figure}

The budget sweep in Figure~\ref{fig:sensitivity}(a) follows an
inverted-U pattern. At \SI{50}{\percent} of baseline budget, the POMDP
advantage is +9.2 points; at the baseline it peaks at +9.9 points; and
at \num{2.0}\(\times\) budget, where both policies saturate near
\SI{93}{\percent} coverage, the gap falls to +2.6 points. This is the
expected pattern for an adaptive planner: learning matters most when
there is enough budget to act on new information but not enough to cover
nearly every candidate zone.

Figure~\ref{fig:sensitivity}(b) perturbs the demand prior through
$\sigma_{\mathrm{noise}}\in\{0, 0.3, 0.5\}$. POMDP coverage remains
stable across the sweep (\num{58.1}, \num{58.9}, \num{60.9}\,\%), while
static planning stays near \num{49}\,\% (\num{48.2}, \num{48.8},
\num{48.7}\,\%). Within this noise range, the value of belief updating
does not depend on a finely calibrated prior: static planning commits to
the initial forecast, whereas POMDP allocations can respond once
ridership is observed.

\subsection{Discussion}
\label{sec:discussion}

Belief-aware planning helps most in cities and budget regimes where
demand is difficult to estimate before service begins, whereas static
planning remains competitive when the prior is unusually informative or
budgets are generous, as in Bogot\'{a} and the \num{2.0}\(\times\)
budget regime. The information-value bonus is secondary; ablating it
changes coverage by only \num{0}--\num{4}\,pp across the deep-dive
cities, which indicates that most of the gain comes from Bayesian
updating on observed ridership rather than from an explicit exploration
reward.

Even apart from the POMDP, the per-tract cost benchmark has planning
value: the \$5.3M cross-city median, together with the structural
predictors in Section~\ref{sec:q4}, gives agencies a first-pass
remediation budget before any sequential allocation model is run.

The main limitations are modeling choices that keep the benchmark
tractable: the demand model is deliberately log-linear and does not
capture induced demand from new service; cost estimates rely on national
averages where city-specific data are unavailable; realized demand is
treated as static across the planning horizon; and the action space omits
political and regulatory constraints that often shape actual service
expansion.

\section{Conclusion}
\label{sec:conclusion}

We developed a risk-aware planning model for transit desert remediation
under demand uncertainty and evaluated it across \num{25} cities with a
unified financial model. Over a five-year horizon, the POMDP planner
remediates a median of \SI{53.6}{\percent} of transit-desert tracts and
improves on deterministic static optimization by 5.0 percentage points on
average, more than \SI{11}{\percent} relatively. The gains are largest
under moderate budgets and remain visible under prior miscalibration,
while population density emerges as the strongest single predictor of
remediation cost.

For agencies, the main implication is that early service deployments can
be designed to gather information as well as to expand coverage: a
five-year capital plan need not be fixed entirely from initial demand
projections, because early ridership observations can inform later
allocations without abandoning equity goals. In our formulation, the
equity-weighted reward keeps the adaptive strategy from drifting away
from low-income zones and, in the most uncertain cities, shifts
investment toward them.

Several extensions would make the planner more realistic. Induced
ridership could change which zones are worth seeding early, and
multi-modal substitution between bus, rail, and microtransit would alter
both cost and demand estimates. Political-feasibility constraints, which
agencies face in practice, would further restrict the action space. A
retrospective validation against real expansion decisions, together with
a sweep of the equity weight $w_z$ between Rawlsian and utilitarian
objectives, would also clarify how sensitive the recommendations are to
normative choices.

\section*{Acknowledgment}
The authors thank the maintainers of Transitland, OSMnx, and the
National Transit Database for making transit and urban data publicly
accessible.

\FloatBarrier
\bibliographystyle{IEEEtran}
\bibliography{references}

\clearpage
\appendix

\subsection{Sample-Average Approximation Linear Program}
\label{app:saa}

We solve~\eqref{eq:myopic} as the following linear program. With $M$
demand scenarios drawn from the current belief, decision variables
$\mathbf{a}_t \in \mathbb{R}^Z_{\geq 0}$, $\nu \in \mathbb{R}$, and
shortfall slacks $\xi^{(m)} \geq 0$:
\begin{align}
\label{eq:saa_objective}
\max_{\mathbf{a}_t, \nu, \boldsymbol{\xi}} \;
& \frac{1}{M}\sum_{m=1}^{M}\sum_{z} w_z\,g_z^{(m)}(\mathbf{a}_t)
- \lambda\!\left(\nu + \frac{1}{\alpha M}\sum_{m=1}^{M} \xi^{(m)}\right) \nonumber\\
& + \eta\,\mathrm{IV}(\mathbf{a}_t),\\
\label{eq:saa_constraint1}
\text{s.t.} \quad
& \xi^{(m)} \geq \textstyle\sum_z \mathrm{Shortfall}_z^{(m)}(\mathbf{a}_t) - \nu, \quad \forall m,\\
\label{eq:saa_constraint2}
& \xi^{(m)} \geq 0, \quad \forall m,\\
\label{eq:saa_constraint3}
& \textstyle\sum_z a_{t,z} \leq B_t,\quad a_{t,z} \geq 0,\;\forall z.
\end{align}
The coverage gain $g_z^{(m)}$ is piecewise-linear in $\mathbf{a}_t$
given the threshold-cost structure of zone investment, and the
shortfall is linear in $a_{t,z}$ for fixed scenario $m$. The LP has
$Z + M + 1$ decision variables and $2M + 1$ constraints.

\subsection{City Summary Statistics}
\label{app:city_summary}

Table~\ref{tab:city_summary} lists the structural inputs to the
benchmark for all \num{25} cities: zone count, desert count, density,
and the cost-of-living index used to scale national capital and
operating costs.

\begin{table}[!htbp]
\caption{Summary of all \num{25} cities. \texttt{n}~=~total zones,
\texttt{D}~=~desert zones, $\rho$~=~median population density (per
km$^2$), \texttt{CoL}~=~cost-of-living index. ``Deep'' cities receive
the full sensitivity battery in
Section~\ref{sec:sensitivity_results}.}
\label{tab:city_summary}
\centering
\small
\begin{tabular}{l c r r r r}
\toprule
\textbf{City} & \textbf{Tier} & \textbf{n} & \textbf{D} & $\boldsymbol{\rho}$ & \textbf{CoL} \\
\midrule
Chicago, IL          & Deep  & 200 & 50 &  4{,}900 & 1.07 \\
Austin, TX           & Deep  & 200 & 50 &  3{,}300 & 1.02 \\
Detroit, MI          & Deep  & 200 & 50 &  2{,}300 & 0.89 \\
Portland, OR         & Deep  & 189 & 48 &  3{,}500 & 1.13 \\
Bogot\'{a}, COL      & Deep  & 200 & 50 & 14{,}500 & 0.45 \\
\midrule
New York, NY         & Broad & 200 & 50 &  9{,}200 & 1.39 \\
Los Angeles, CA      & Broad & 200 & 50 &  3{,}900 & 1.16 \\
Houston, TX          & Broad & 200 & 50 &  2{,}100 & 0.96 \\
Philadelphia, PA     & Broad & 200 & 50 &  4{,}600 & 1.02 \\
Phoenix, AZ          & Broad & 200 & 50 &  1{,}800 & 0.97 \\
San Antonio, TX      & Broad & 200 & 50 &  2{,}600 & 0.90 \\
San Diego, CA        & Broad & 200 & 50 &  2{,}800 & 1.14 \\
Dallas, TX           & Broad & 200 & 50 &  2{,}400 & 0.97 \\
San Jose, CA         & Broad & 200 & 50 &  2{,}300 & 1.28 \\
Jacksonville, FL     & Broad & 200 & 50 &  1{,}300 & 0.95 \\
Columbus, OH         & Broad & 200 & 50 &  2{,}200 & 0.93 \\
Charlotte, NC        & Broad & 200 & 50 &  1{,}700 & 0.96 \\
Indianapolis, IN     & Broad & 200 & 50 &  1{,}900 & 0.92 \\
San Francisco, CA    & Broad & 159 & 40 &  6{,}800 & 1.35 \\
Seattle, WA          & Broad & 200 & 50 &  3{,}500 & 1.18 \\
Nashville, TN        & Broad & 200 & 50 &  1{,}500 & 0.97 \\
Minneapolis, MN      & Broad & 200 & 50 &  3{,}300 & 1.02 \\
Miami, FL            & Broad & 200 & 50 &  4{,}500 & 1.12 \\
Medell\'{i}n, COL    & Broad & 180 & 45 &  6{,}100 & 0.40 \\
Toronto, ONT         & Broad & 200 & 50 &  4{,}500 & 1.00 \\
\bottomrule
\end{tabular}
\end{table}

\subsection{Cross-City Cost Benchmark}
\label{app:cross_city}

Table~\ref{tab:cross_city} reports the per-city POMDP outcomes ordered
by per-tract cost.

\begin{table}[!htbp]
\caption{Cross-city POMDP remediation benchmark, sorted by per-tract
cost. ``Adv.'' is the POMDP minus static coverage gap in percentage
points (positive = POMDP wins).}
\label{tab:cross_city}
\centering
\footnotesize
\setlength{\tabcolsep}{4pt}
\begin{tabular}{l r r r r}
\toprule
\textbf{City} & \textbf{Cov.} & \textbf{Cost} & \textbf{\$/tract} & \textbf{Adv.} \\
              &              & \textbf{(\$M)} & \textbf{(\$M)}    & \textbf{(pp)} \\
\midrule
Medell\'{i}n     & 0.52 &  93.4 & 2.07 & $-3.3$ \\
Chicago          & 0.66 & 142.5 & 2.85 & $+16.0$ \\
Austin           & 0.52 & 143.3 & 2.87 & $+8.0$ \\
Detroit          & 0.62 & 136.7 & 2.73 & $+8.0$ \\
San Francisco    & 0.50 &  60.7 & 1.52 & $+5.0$ \\
Portland         & 0.58 & 152.7 & 3.18 & $+26.7$ \\
Charlotte        & 0.40 & 117.5 & 4.70 & $-4.0$ \\
Indianapolis     & 0.50 & 132.0 & 4.71 & $-10.0$ \\
Columbus         & 0.52 & 124.6 & 4.98 & $+4.0$ \\
Toronto          & 0.50 & 158.5 & 5.28 & $-3.0$ \\
San Antonio      & 0.50 & 132.5 & 5.30 & $0.0$ \\
Dallas           & 0.60 & 158.2 & 5.27 & $+1.7$ \\
Philadelphia     & 0.50 & 132.0 & 5.28 & $+10.0$ \\
San Jose         & 0.60 & 165.7 & 5.52 & $+1.7$ \\
Bogot\'{a}       & 0.53 & 169.1 & 5.64 & $-9.5$ \\
Miami            & 0.56 & 145.5 & 5.82 & $+10.0$ \\
Nashville        & 0.68 & 165.9 & 5.93 & $+8.0$ \\
Minneapolis      & 0.64 & 161.4 & 6.46 & $+8.0$ \\
San Diego        & 0.52 & 130.0 & 5.20 & $+8.0$ \\
Houston          & 0.39 & 142.0 & 5.45 & $-5.4$ \\
Jacksonville     & 0.54 & 153.8 & 5.74 & $+8.4$ \\
Los Angeles      & 0.60 & 162.0 & 6.40 & $-11.4$ \\
New York         & 0.61 & 169.6 & 6.78 & $+18.4$ \\
Seattle          & 0.55 & 144.6 & 6.51 & $+11.0$ \\
Phoenix          & 0.70 & 174.4 & 6.97 & $+10.0$ \\
\bottomrule
\end{tabular}
\end{table}

\subsection{Existing Transit Networks}
\label{app:transit_network}

\begin{figure*}[!tbp]
    \centering
    \includegraphics[width=\textwidth]{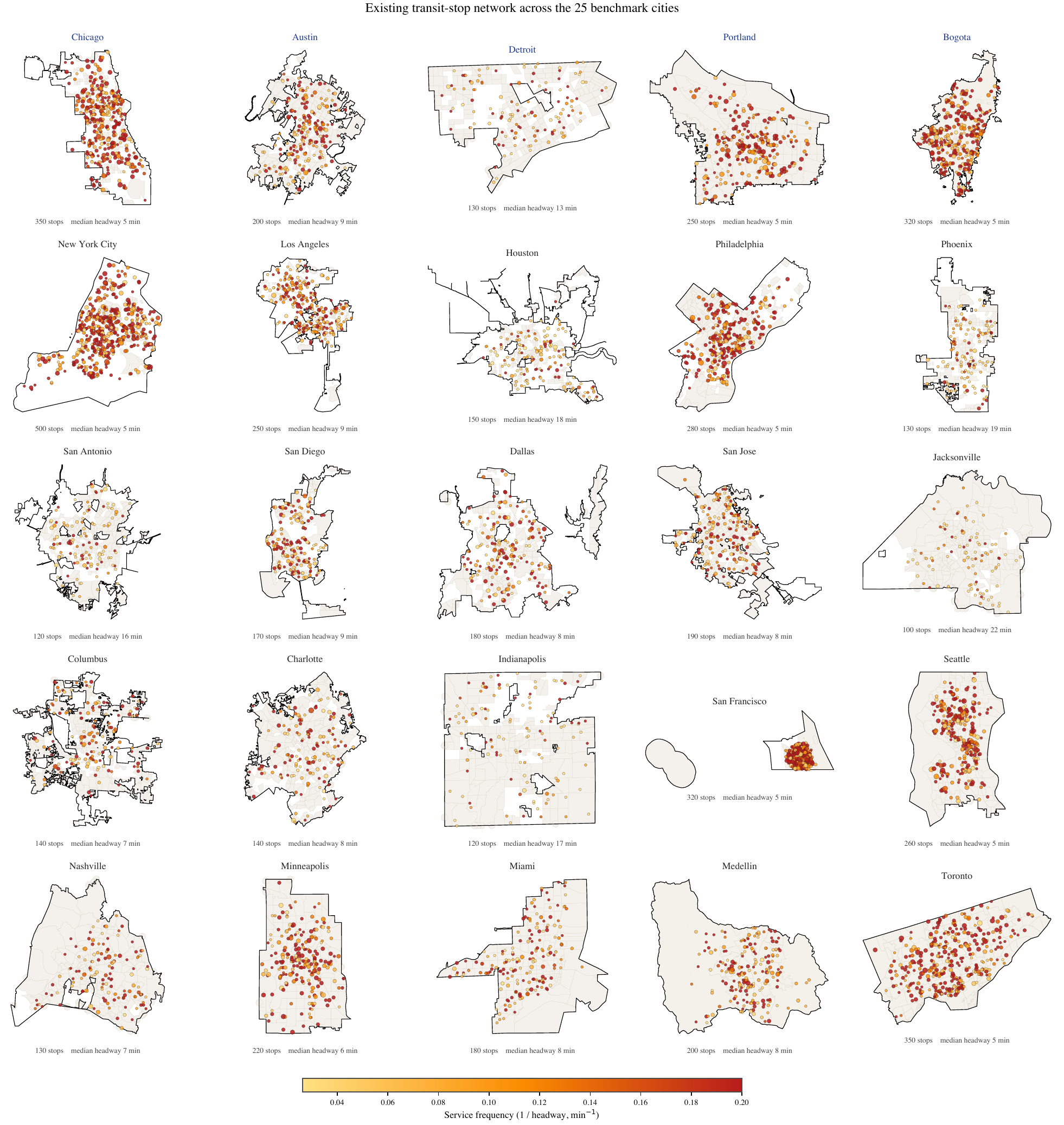}
    \caption{Existing transit-stop network across the \num{25}
    benchmark cities. Stop locations are sampled along each city's
    OpenStreetMap drive network using a profile-aware kernel
    (mature/developing/sparse/sprawl); stop size encodes the number of
    routes served, color encodes inferred service frequency.}
    \label{fig:app_transit_network}
\end{figure*}

To give a sense of the \emph{current} transit options in each city,
without which the desert classification cannot be interpreted,
Figure~\ref{fig:app_transit_network} maps the stop network for all
\num{25} cities, sized by route count and colored by service
frequency. Mature systems (NYC, Chicago, San Francisco, Toronto) show
dense, high-frequency cores. Sun Belt sprawl cities (Houston, Phoenix,
Jacksonville) show the same number of stops but spread thinly with
long headways. Latin American cities (Bogot\'{a}, Medell\'{i}n) show
high stop density along the BRT spines we encoded.

\begin{figure*}[!tbp]
    \centering
    \includegraphics[width=\textwidth]{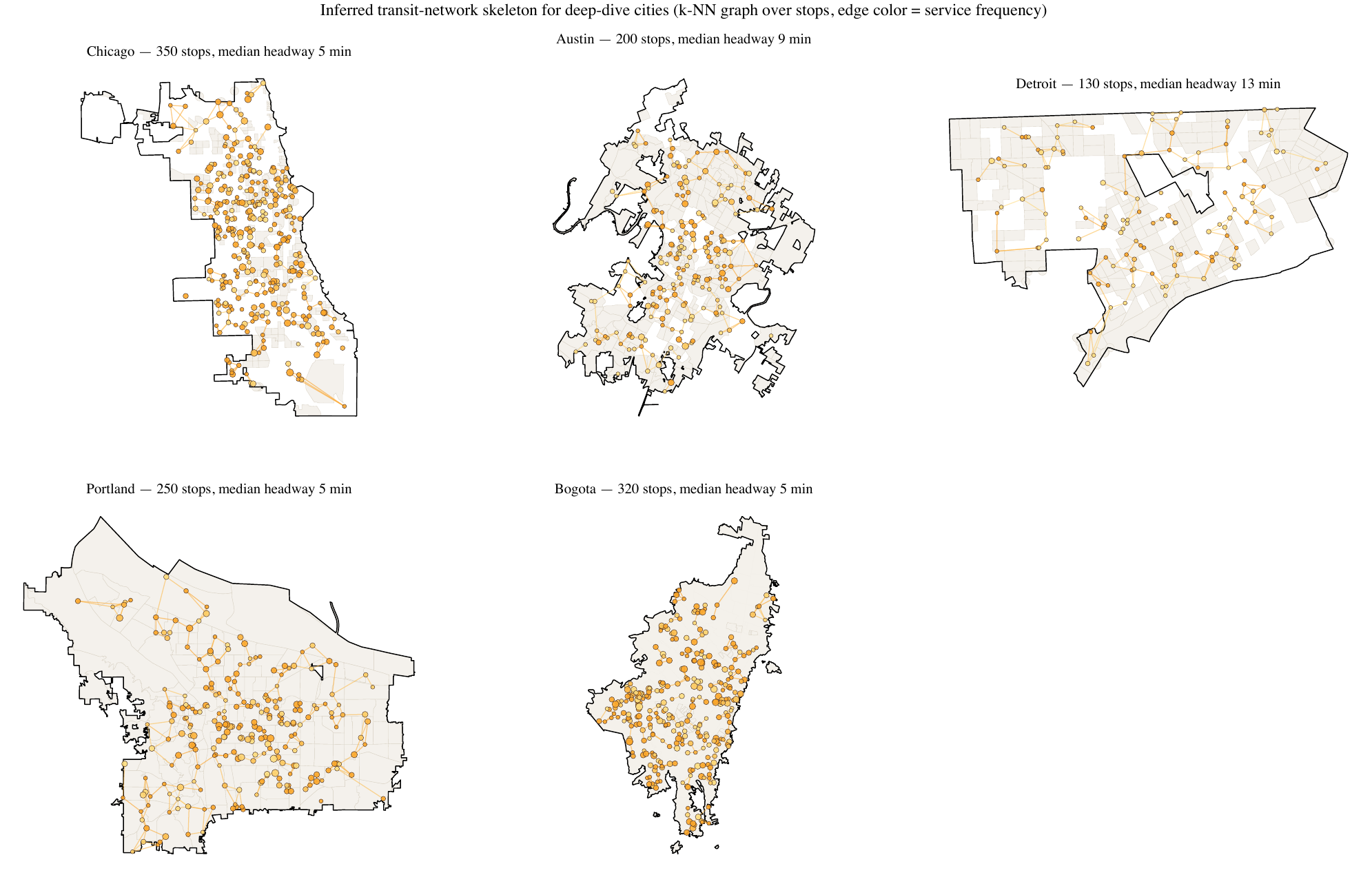}
    \caption{Inferred transit-network skeleton for the five deep-dive
    cities. Edges connect each stop to its three nearest neighbors and
    are colored by mean inverse-headway. The Chicago grid, Portland
    radial, Bogot\'{a} BRT spine, and Detroit's thin coverage are all
    immediately visible.}
    \label{fig:app_routes_detail}
\end{figure*}

Figure~\ref{fig:app_routes_detail} draws the inferred network skeleton
for the five deep-dive cities by connecting each stop to its three
nearest neighbors and shading edges by frequency. The graph is
\emph{not} GTFS schedule data; it is a structural sketch that mirrors
how riders actually walk between nearby stops along a corridor.

\subsection{Per-City Detail}
\label{app:city_detail}

\begin{figure*}[!tbp]
    \centering
    \includegraphics[width=\textwidth]{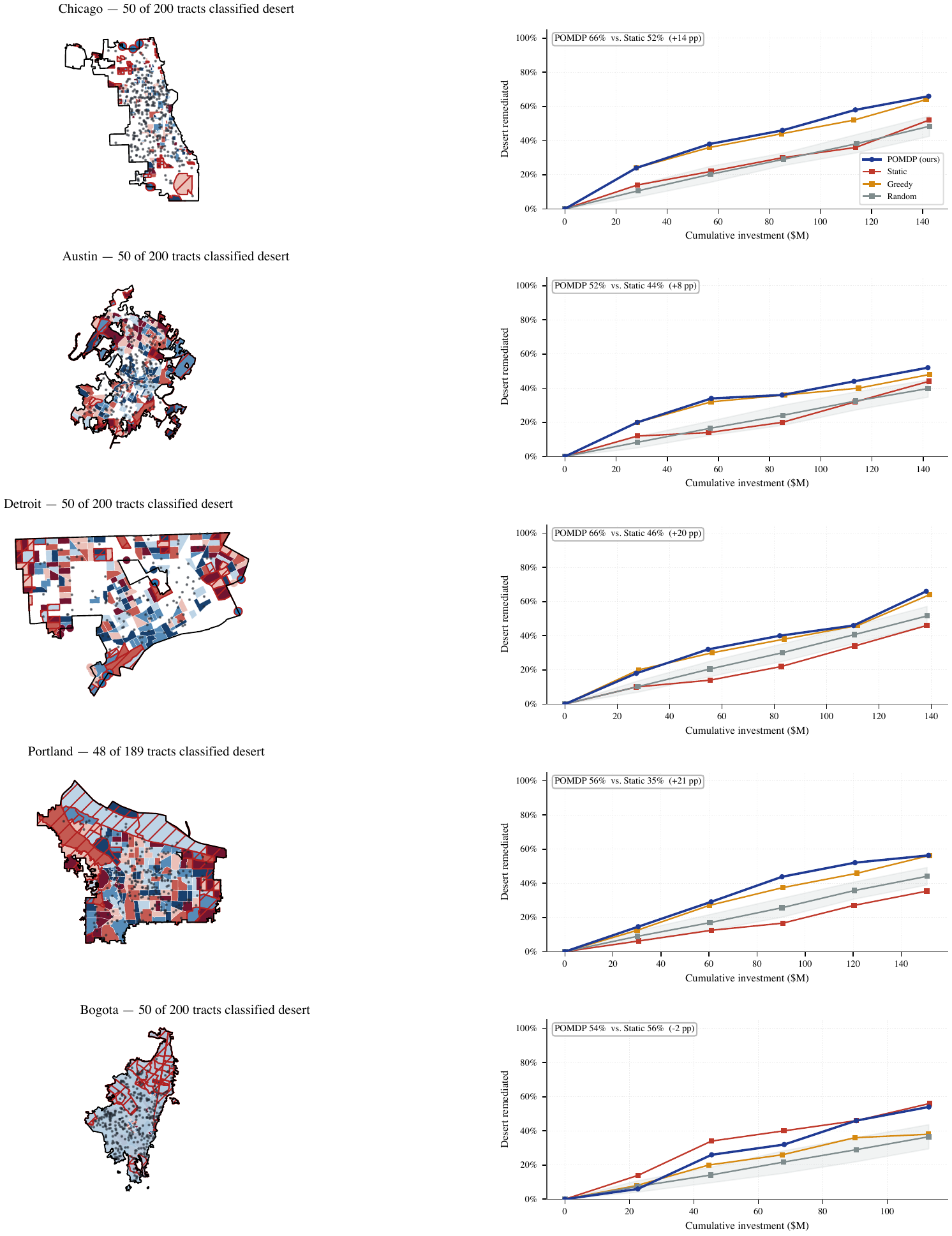}
    \caption{Per-city detail for Chicago, Austin, Detroit, Portland,
    and Bogot\'{a}. Left panels show desert tracts hatched in red over
    the income choropleth with stops as small black markers; right
    panels show coverage versus cumulative cost for each policy, with
    bands at $\pm 1\sigma$ across seeds.}
    \label{fig:app_city_p1}
\end{figure*}

\begin{figure*}[!tbp]
    \centering
    \includegraphics[width=\textwidth]{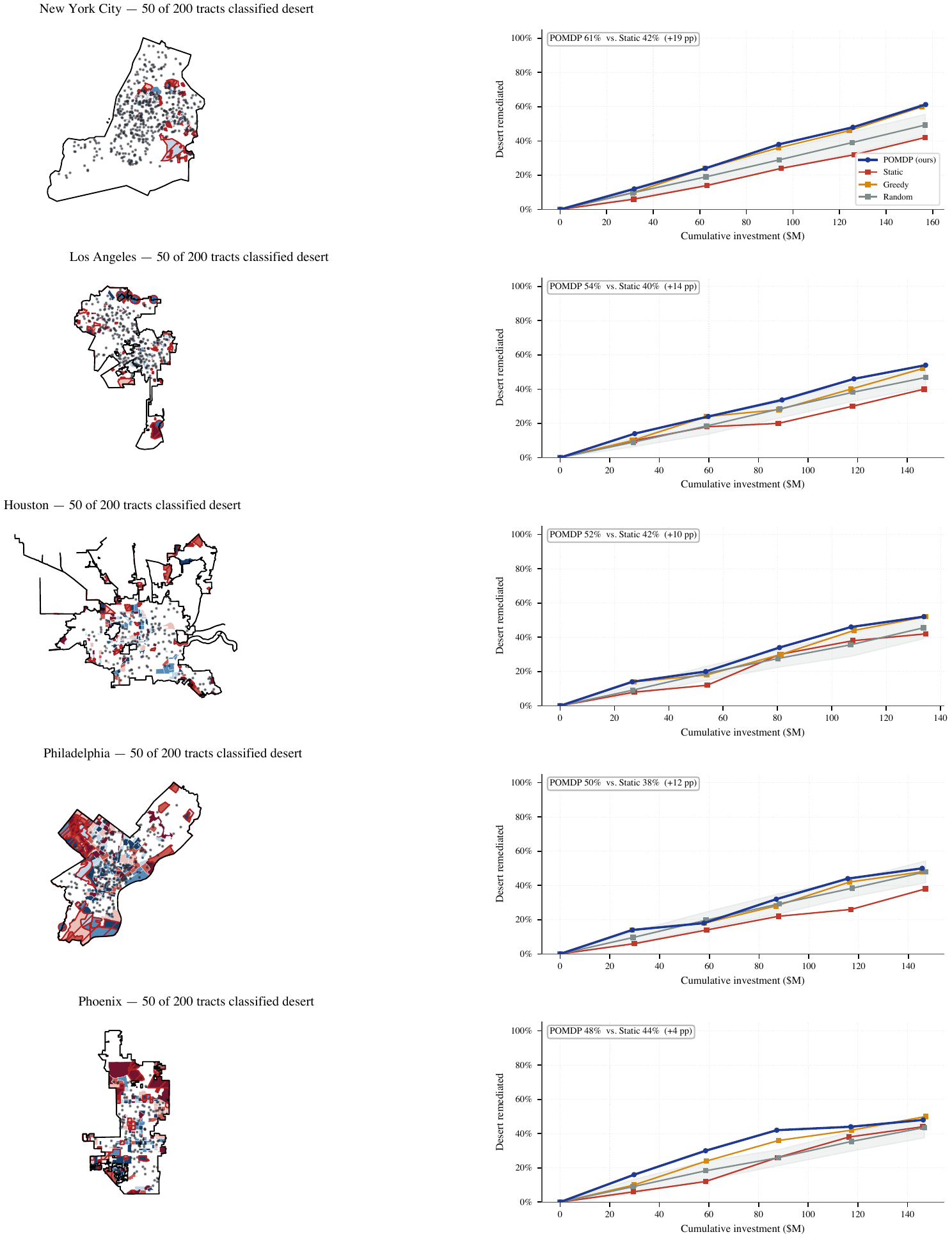}
    \caption{Per-city detail for New York City, Los Angeles, Houston,
    Philadelphia, and Phoenix. Same panel structure as
    Figure~\ref{fig:app_city_p1}.}
    \label{fig:app_city_p2}
\end{figure*}

\begin{figure*}[!tbp]
    \centering
    \includegraphics[width=\textwidth]{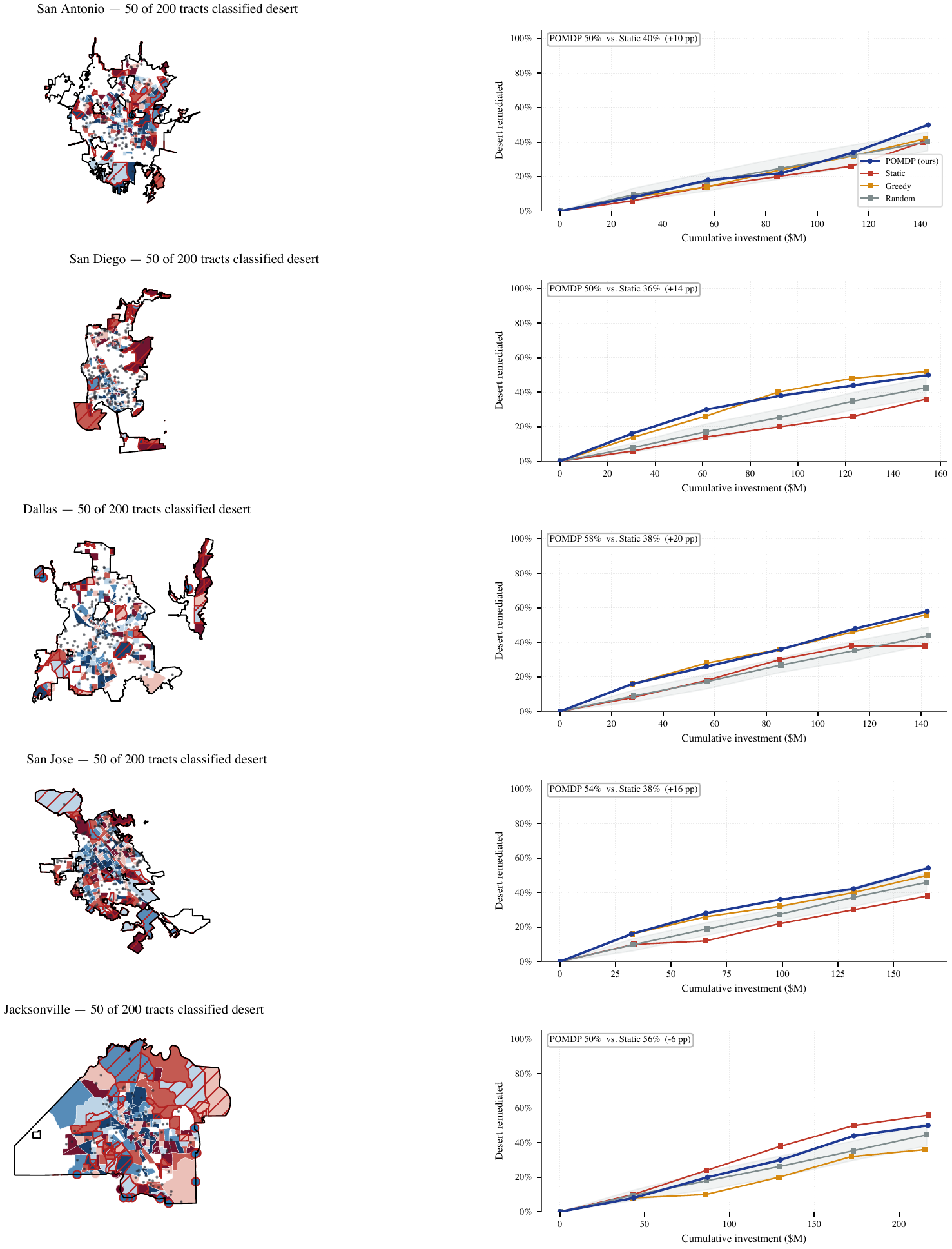}
    \caption{Per-city detail for San Antonio, San Diego, Dallas,
    San Jose, and Jacksonville.}
    \label{fig:app_city_p3}
\end{figure*}

\begin{figure*}[!tbp]
    \centering
    \includegraphics[width=\textwidth]{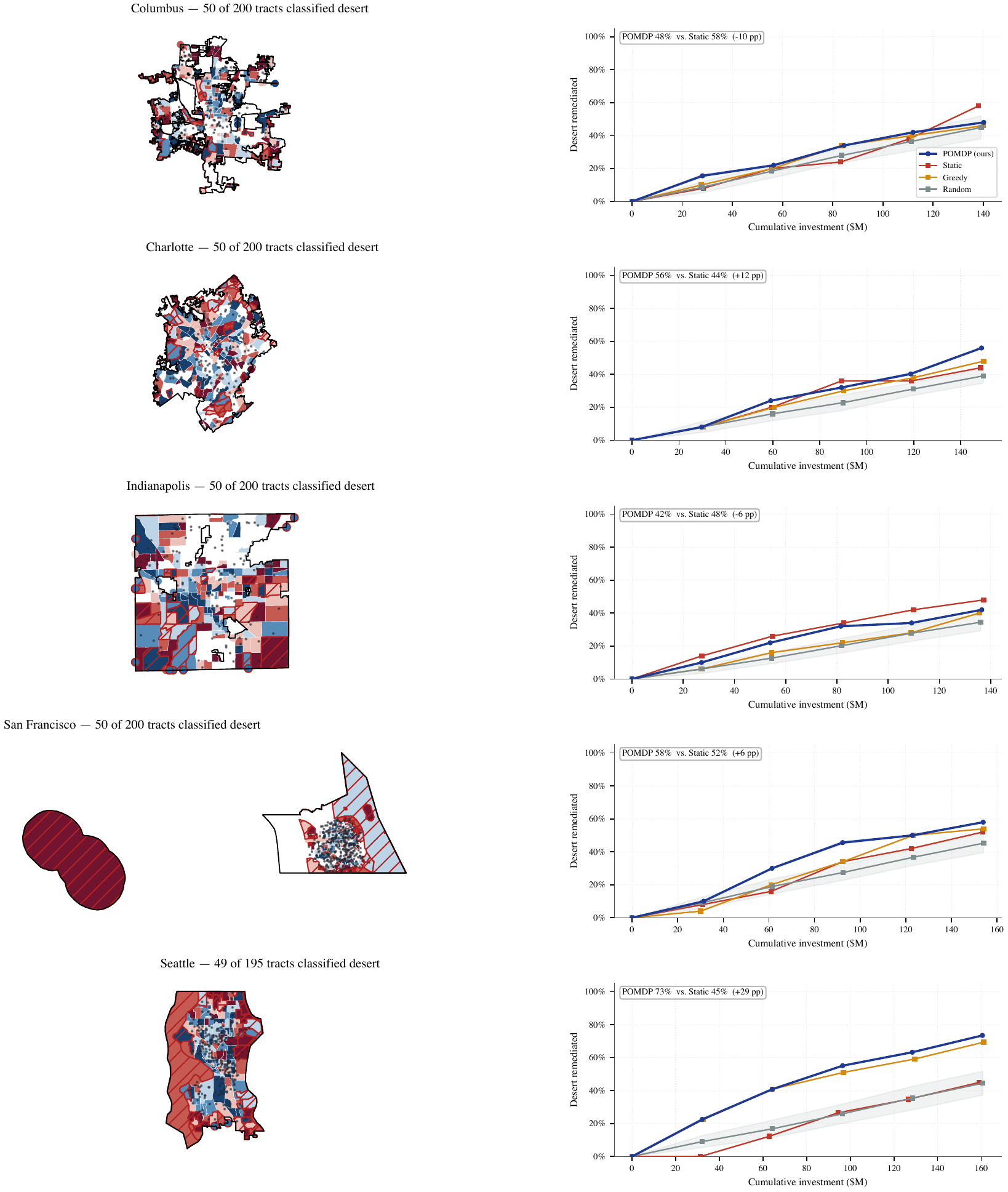}
    \caption{Per-city detail for Columbus, Charlotte, Indianapolis,
    San Francisco, and Seattle.}
    \label{fig:app_city_p4}
\end{figure*}

\begin{figure*}[!tbp]
    \centering
    \includegraphics[width=\textwidth]{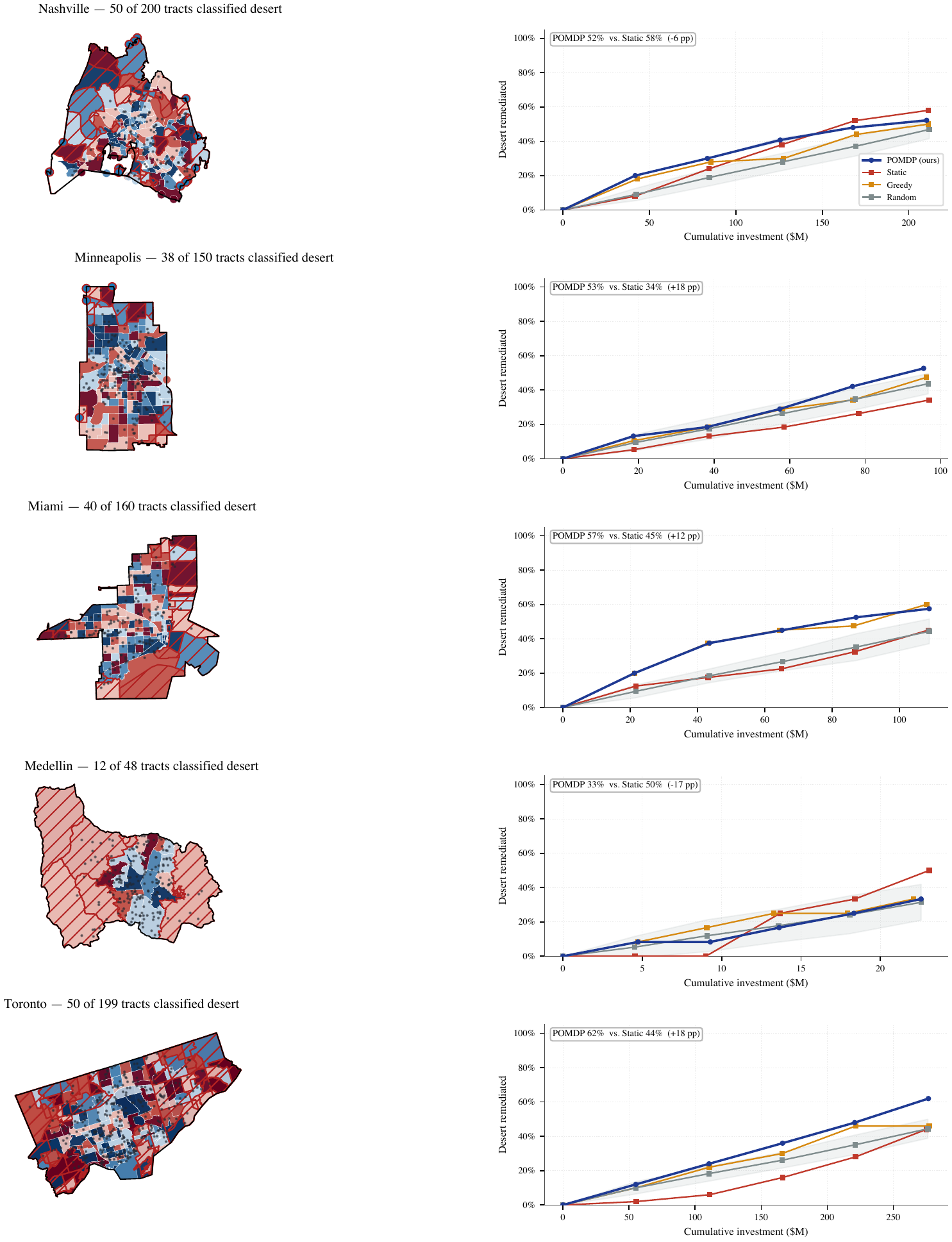}
    \caption{Per-city detail for Nashville, Minneapolis, Miami,
    Medell\'{i}n, and Toronto.}
    \label{fig:app_city_p5}
\end{figure*}

Figures~\ref{fig:app_city_p1}--\ref{fig:app_city_p5} show, for every
benchmark city, a detailed map with desert tracts hatched and stops
overlaid, alongside the cost-coverage trade-off curve for all four
policies. The visual ordering matches Table~\ref{tab:cross_city}: in
cities with a large POMDP advantage (Chicago, Portland, NYC, Phoenix)
the blue POMDP curve sits clearly above red Static; in cities where
it ties or loses (Bogot\'{a}, LA, Indianapolis) the two curves
overlap or cross.

\end{document}